\documentclass[twocolumn,english,aps,prl,reprint,groupedaddress,notitlepage,nobibnotes,nofootinbib,preprintnumbers,showpacs]{revtex4-1}
\pdfoutput=1
\usepackage{lmodern}

\usepackage[T1]{fontenc}
\usepackage[latin9]{inputenc}
\usepackage{geometry}
\usepackage{color}
\usepackage{babel}
\usepackage{mathtools}
\usepackage{amsmath}
\usepackage{amssymb}
\usepackage{graphicx}
\usepackage{esint}
\usepackage[unicode=true,pdfusetitle,
 bookmarks=true,bookmarksnumbered=false,bookmarksopen=false,
 breaklinks=false,pdfborder={0 0 1},backref=false,colorlinks=true]
 {hyperref}
\hypersetup{
 citecolor=black,linkcolor=black,urlcolor=black}
\makeatletter

 \@ifundefined{textcolor}{}
 {
   \definecolor{BLACK}{gray}{0}
   \definecolor{WHITE}{gray}{1}
   \definecolor{RED}{rgb}{1,0,0}
   \definecolor{GREEN}{rgb}{0,1,0}
   \definecolor{BLUE}{rgb}{0,0,1}
   \definecolor{CYAN}{cmyk}{1,0,0,0}
   \definecolor{MAGENTA}{cmyk}{0,1,0,0}
   \definecolor{YELLOW}{cmyk}{0,0,1,0}
 }
\usepackage{babel}
\makeatletter
\def\simgt{\mathrel{\lower2.5pt\vbox{\lineskip=0pt\baselineskip=0pt
           \hbox{$>$}\hbox{$\sim$}}}}
\def\simlt{\mathrel{\lower2.5pt\vbox{\lineskip=0pt\baselineskip=0pt
           \hbox{$<$}\hbox{$\sim$}}}}
\makeatother

\newcommand{\be}{\begin{equation}}
\newcommand{\ee}{\end{equation}}
\newcommand{\bea}{\begin{eqnarray}}
\newcommand{\eea}{\end{eqnarray}}
\newcommand{\Ref}[1]{Ref.~\cite{#1}}

\newcommand{\Eq}[1]{Eq.~\eqref{#1}}
\newcommand{\Eqs}[2]{Eqs.~\eqref{#1} and \eqref{#2}}

\begin{document}

\preprint{\hbox{CALT-TH-2016-018} }

\title{Positivity of Curvature-Squared Corrections in Gravity}

\author{Clifford Cheung and Grant N. Remmen}
\affiliation{Walter Burke Institute for Theoretical Physics \\California Institute of Technology, Pasadena, CA 91125}
\date{\today}
\email{clifford.cheung@caltech.edu, gremmen@theory.caltech.edu}
\pacs{}
\begin{abstract}
We study the Gauss-Bonnet (GB) term as the leading higher-curvature correction to pure Einstein gravity.  Assuming a tree-level ultraviolet completion free of ghosts or tachyons, we prove that the GB term has a nonnegative coefficient in dimensions greater than four. Our result follows from unitarity of the spectral representation for a general ultraviolet completion of the GB term. 
\end{abstract}

\maketitle

\section{Introduction\label{sec:intro}}

Effective field theory lore states that, in constructing a Lagrangian, one should include all operators allowed by symmetry and power counting with arbitrary coefficients.  Naively, this implies an immense freedom for low-energy model-building.  However, not all quantum effective field theories are created equal: some are compatible with ultraviolet completion, while others reside in the so-called swampland \cite{IRUV,Vafa:2005ui,Ooguri:2006in}, impervious to string-theoretic completion or, worse, any completion conforming to the usual axioms of quantum field theory.  

An ongoing effort has been undertaken to demarcate the boundaries of healthy effective field theories, with constraints derived from both top-down and bottom-up reasoning.  An iconic example of the former is the weak gravity conjecture \cite{WGC}, which was deduced from string-theoretic examples and black hole thought experiments.  In the latter approach, one conceives bounds purely within the logic of low-energy effective theory,  e.g., from considerations of causality, unitarity, and locality/analyticity for long-distance observables such as scattering amplitudes and particle trajectories \cite{IRUV,Pham:1985cr,Distler:2006if,Nicolis:2009qm,
Rattazzi,Komargodski:2011vj,Brando2,Baumann:2015nta,
Brando,Jenkins:2006ia,Dvali:2012zc,Aharonov:1969vu,Maldacena3pt,
IRWGC,GravityBounds,MassiveGravity}.  

In this paper, we derive a simple bound on curvature-squared corrections to Einstein gravity. Taking a low-energy perspective, we study gravity as an effective field theory described by the Einstein-Hilbert action,\footnote{We use mostly $+$ signature for the metric, adopt sign conventions $R_{\mu\nu} = R^\rho_{\;\;\mu\rho\nu}$ and $R^\mu_{\;\;\nu\rho\sigma} = \partial_\rho \Gamma^\mu_{\;\;\nu\sigma}+\cdots$ for the curvature tensor, and define $\kappa = \sqrt{8 \pi G}$.} $S=\int \mathrm{d}^D x \sqrt{-g} \, R/2\kappa^2$, whose  higher-curvature corrections {\it a priori} include $R_{\mu\nu\rho\sigma}R^{\mu\nu\rho\sigma}$, $R_{\mu\nu}R^{\mu\nu}$, and $R^2$.  However, the usual invariance under field redefinitions implies that leading corrections in the derivative expansion are defined only up to equations of motion, so those operators involving $R$ and $R_{\mu\nu}$ can be discarded.  Hence, the only nontrivial leading correction to pure Einstein gravity is effectively $R_{\mu\nu\rho\sigma}R^{\mu\nu\rho\sigma}$, which up to equations of motion is equivalent to the Gauss-Bonnet (GB) term
\be
\begin{aligned}
\Delta S\!=\!\! \int \!\! {\rm d}^D x \sqrt{-g}\, \lambda\! \left(R_{\mu\nu\rho\sigma}R^{\mu\nu\rho\sigma}\! - 4R_{\mu\nu}R^{\mu\nu} + R^2\right)\!.\label{eq:action}
\end{aligned}
\ee
The GB term is a total derivative in $D=4$, so we take $D>4$ throughout. The GB term is ghost-free~\cite{Zwiebach} and is ubiquitous in string-theoretic completions of gravity. 

The coupling constant $\lambda$ is an important low-energy probe of the ultraviolet completion of general relativity.  The sign of $\lambda$ is also of particular interest from holographic considerations, being related to the viscosity-to-entropy ratio of the dual conformal field theory (see \Ref{PavelKats} and refs. therein). More importantly, $\lambda \geq0$ appears to be a generic prediction of string theory: $\lambda=0$ in type II superstring theory \cite{GrossWitten}, while $\lambda>0$ for the bosonic~\cite{Zwiebach}, heterotic~\cite{GrossSloan}, and type I~\cite{Tseytlin} string. 

Here, we explore theories in which the GB term is generated by weakly-coupled dynamics below the Planck scale, corresponding to large $\lambda$ in natural units.  Furthermore, we assume that ``primordial'' contributions to the GB term---i.e., contributions present in the ultraviolet but unaccompanied by new states---are subdominant.  This assumption is reasonable because a primordial GB term will violate unitarity below the Planck scale.  In addition, \Ref{Maldacena3pt} demonstrated how a primordial GB term violates causality unless new states are introduced. Moreover, it can incur potential violations of analyticity~\cite{GravityBounds} and the second law of black hole thermodynamics~\cite{Sarkar:2010xp}.   All of these issues strongly motivate consideration of a GB term generated dominantly by weakly-coupled ultraviolet dynamics.

Within these assumptions, we will prove that $\lambda \geq 0$ for any unitary tree-level ultraviolet completion of the GB term.  To do so, we first enumerate interactions that couple gravitons to massive states in order to generate the GB term at tree level. We then introduce a general spectral representation for the two-point function for these massive degrees of freedom.  Finally, we show how unitarity of the spectral representation fixes the sign of the curvature-squared operator coefficient in the gravitational effective theory.

\section{Coupling to Massive States\label{sec:UV}}

In this section, we study the structure of weakly-coupled ultraviolet dynamics that generates curvature-squared corrections to gravity at low energies.  As noted earlier, we can freely substitute the tree-level equations of motion---i.e., Einstein's equations---into the leading curvature corrections in \Eq{eq:action}.   In practice, this means that the GB term is, at leading order in the derivative expansion, equivalent to the Riemann-squared operator and the Weyl-squared operator,
\be 
\begin{aligned}
C_{\mu\nu\rho\sigma}C^{\mu\nu\rho\sigma} =& R_{\mu\nu\rho\sigma}R^{\mu\nu\rho\sigma} -\frac{4}{D-2}R_{\mu\nu}R^{\mu\nu} \\&+ \frac{2}{(D-1)(D-2)}R^2,
\end{aligned}
\ee
where the Weyl tensor is
\be
\begin{aligned}
C_{\mu\nu\rho\sigma} =& R_{\mu\nu\rho\sigma} - \frac{1}{D-2}\left(g_{\mu[\rho}R_{\sigma]\nu}  - g_{\nu[\rho}R_{\sigma]\mu}\right) \\&+ \frac{1}{(D-1)(D-2)}Rg_{\mu[\rho}g_{\sigma]\nu}
\end{aligned}
\ee
and square brackets on indices denote antisymmetrization without normalization, i.e., $A_{[\mu\nu]}= A_{\mu\nu} - A_{\nu\mu}$.  In the presence of massless matter or gauge fields, this equivalence holds modulo additional interactions involving the stress-energy tensor.

This all implies that the low-energy coefficients of the GB term, the Riemann-squared term, and the Weyl-squared term are equal.  For technical simplicity, we therefore recast the action as
\be
S = \int {\rm d}^D x \sqrt{-g}\left( \frac{R}{2\kappa^2} + \lambda C_{\mu\nu\rho\sigma} C^{\mu\nu\rho\sigma}  \right)\label{eq:actionWeyl}
\ee
using the freedom of equations of motion.  Let us note that the above action applies to a low-energy theory comprised purely of massless gravitons.  If there are additional spectator massless matter fields or gauge fields, there will be additional terms involving the stress-energy tensor that do not affect our arguments.

Throughout our analysis, we assume a weakly-coupled ultraviolet completion of gravity.  In turn, this assumption implies that high-energy graviton scattering is unitarized by tree-level exchanges of heavy states.
 The reason for this is as follows. In any theory that is weakly coupled from the ultraviolet to the infrared, there is, by definition, a well-defined $\hbar$ expansion at all scales.  Crucially, in general relativity, diffeomorphism symmetry relates the kinetic term for the graviton to its interactions within the Einstein-Hilbert term.  Since the former is manifestly an $\mathcal{O}(1/\hbar)$ tree-level effect, then so, too, is the latter, which means that it can only be unitarized by tree-level exchanges.  
 
A similar line of reasoning applies to the nonlinear sigma model, which is why unitarization of pion scattering at weak coupling can only be achieved via tree-level Higgs exchange. More generally, while the weak coupling assumption could potentially be relaxed through an accounting of loop corrections as in \Ref{Distler:2006if}, such an approach would apply to the derivation of positivity bounds via scattering amplitudes and analytic dispersion relations (e.g., \Ref{GravityBounds}), as opposed to the unitarity-based methods of the present work.

In contrast to the leading-order gravity action, operators like the GB term are separately diffeomorphism invariant and are not directly connected to the Einstein-Hilbert term via symmetry.  Hence, even at weak coupling, the GB operator can be ultraviolet-completed at tree or loop level.  An analogous statement is true for Euler-Heisenberg higher-dimension operators in gauge theory: since they are not connected directly to the gauge kinetic term, they can arise from tree-level exchange or at loop order.

Nevertheless, since high-energy graviton scattering is unitarized at tree level, it is well motivated to focus on tree-level ultraviolet completions of the GB term.  Indeed, this is how the GB term arises in the low-energy gravitational effective actions of string theories.  Thus, from here on we assume that \Eq{eq:action} arises from the exchange of heavy states at tree level.

Next, let us systematically enumerate all possible ultraviolet-completing dynamics for the GB term.  Denoting a heavy state by $\chi$, we must identify all diffeomorphism-invariant couplings between $\chi$ and gravitons.  These interactions could involve one, two, or more powers of $\chi$, which we now consider.

For interactions that are linear in $\chi$, any derivatives on $\chi$ can always be shuffled onto the gravitons via integration by parts.  Since $\chi$ is like a matter field, it by construction transforms as a tensor and thus necessarily couples to some combination of gravitons that also transforms as a tensor.\footnote{By tensor, we simply mean an object that transforms covariantly under nonlinear coordinate transformations.  Since the metric $g_{\mu\nu}$ is a tensor, it is convenient to parameterize all dependence of the graviton through $g_{\mu\nu}$, its  associated curvature tensors, and covariant derivatives $\nabla_\mu$.} If this tensor of gravitons has no derivatives, then in the flat-space limit $\chi$ appears as a tadpole in the Lagrangian, so the corresponding term is eliminated once we expand around the proper vacuum.  On the other hand, if this tensor has exactly one derivative, then the resulting operator must be a total derivative since the metric is covariantly constant.  Finally, if this tensor has two derivatives, then it has mass dimension two and thus just the right power counting to induce a curvature-squared operator.  Indeed, any more derivatives will generate operators of higher order than curvature-squared in the derivative expansion.  

The only possible tensors of mass dimension two constructed from the metric are the Riemann tensor and its contractions~\cite{Weinberg:1972kfs}. Hence, any graviton interactions that are linear in $\chi$ must take the form
\be
y\,C_{\mu\nu\rho\sigma}\chi^{\mu\nu\rho\sigma},\label{eq:operator}
\ee
where $\chi_{\mu\nu\rho\sigma}$ is a field representing all the massive states that generate the GB term and $y$ is a coupling constant.  Analogous operators involving $R_{\mu\nu}$ and $R$ can be discarded by equations of motion.  

Without loss of generality, we can take $\chi_{\mu\nu\rho\sigma}$ in \Eq{eq:operator} to possess all of the index properties of the Weyl tensor, namely, the requisite (anti-)symmetries, the first Bianchi identity, and on-shell tracelessness. Any components of $\chi_{\mu\nu\rho\sigma}$ that violate these symmetry properties are automatically projected out by the Weyl tensor in \Eq{eq:operator}.

Note that \Eq{eq:operator} induces {\it mixing} between the graviton and the heavy state.  However, since this preserves diffeomorphism invariance, the resulting massless eigenstate should still be interpreted as the massless graviton.

On the other hand, interactions that are quadratic in $\chi$ will automatically produce new heavy states in pairs.  To generate an effective operator involving only gravitons, we can close the loop of heavy states, but this interaction goes beyond tree level and is thus suppressed at weak coupling.  An important exception to this occurs if $\chi$ mixes with the graviton, in which case we must introduce \Eq{eq:operator} anyway.  Similar arguments apply for interactions with higher powers of $\chi$, but the final result is the same: any weakly-coupled ultraviolet completion of the GB term will involve the operator in \Eq{eq:operator}.

\section{Spectrum of Massive States\label{sec:spectral}}
\vspace{-0.3em}
Next, we construct a general K{\"a}ll{\'e}n-Lehmann spectral representation \cite{Kallen,Lehmann} for the heavy states $\chi$ following the analysis of Refs.~\cite{IRWGC,Dvali:2012zc,Jenkins:2006ia}.  By expanding the metric $g_{\mu\nu}$ around a flat background $\eta_{\mu\nu}$, we can represent the $\chi$ two-point function in $D$ dimensions as
\begin{widetext}
\be
\langle\chi_{\mu\nu\rho\sigma}(k)\chi_{\alpha\beta\gamma\delta}(k^\prime)\rangle=i\delta^{D}(k+k^\prime)\int_0^\infty{\rm d}\mu^2
\frac{\rho(\mu^2)}{-k^2-\mu^2+i\epsilon}
\Pi_{\mu\nu\rho\sigma\alpha\beta\gamma\delta},\label{eq:spectralrep}
\ee
\end{widetext}
where $k^2$ is contracted with the flat metric.  Here, $\Pi_{\mu\nu\rho\sigma\alpha\beta\gamma\delta}$ is the propagator numerator for $\chi_{\mu\nu\rho\sigma}$ and $\rho(\mu^2)$ is the spectral density encoding arbitrary ultraviolet dynamics in terms of a distribution of poles corresponding to each massive state.  Since we are working at tree level, $\rho(\mu^2)$ is just a sum over delta functions, so the spectral representation is merely a simple way to package a set of resonances.

The absence of tachyons implies that $\mu^2\geq 0$. As we will soon see, the propagator numerator $\Pi_{\mu\nu\rho\sigma\alpha\beta\gamma\delta}$ is highly constrained by its symmetries and unitarity.  In turn,  $\rho(\mu^2)\geq 0$ is required if  the theory is to be ghost-free \cite{Kallen,Lehmann}. The fact that the spectrum is gapped implies regularity of the two-point function as $k \rightarrow 0$, so the spectral density should vanish as $\mu^2 \rightarrow 0$. 

Unitarity requires that the on-shell propagator numerator be a sum over the tensor product of the physical polarizations \cite{Schwartz}. 
That is, when the on-shell condition $k^2 =-\mu^2$ is satisfied, the propagator numerator is
\be \Pi_{\mu\nu\rho\sigma\alpha\beta\gamma\delta}
=\sum_{i}\varepsilon_{i\mu\nu\rho\sigma}\varepsilon_{i\alpha\beta\gamma\delta}^{*},\label{eq:completeness}
\ee
where $\varepsilon_{i\mu\nu\rho\sigma}$ are the physical polarization states of $\chi_{\mu\nu\rho\sigma}$, indexed by $i$ and normalized so that $\varepsilon_{i\mu\nu\rho\sigma}\varepsilon_{j}^{*\mu\nu\rho\sigma}=\delta_{ij}$. 
By definition, the polarization tensors transform in representations of the $SO(D-1)$ little group for the massive state $\chi_{\mu\nu\rho\sigma}$.  Consequently, the polarizations must reside in the subspace transverse to the momentum of $\chi_{\mu\nu\rho\sigma}$.  From \Eq{eq:completeness}, this implies the transversality condition for on-shell $k_\mu$,
\be 
k^\mu \Pi_{\mu\nu\rho\sigma\alpha\beta\gamma\delta} = 0\label{eq:transverse}
\ee  
and similarly for all other contractions.

Note that $\chi_{\mu\nu\rho\sigma}$ is not a canonical spin-four state \cite{Weinberg1,Weinberg3,Fronsdal,arbitraryspin} since it is not fully symmetric. Rather, as we noted in the previous section, $\chi_{\mu\nu\rho\sigma}$ can without loss of generality be taken to have the index properties of the Weyl tensor, which are then inherited by the corresponding polarizations as well as the propagator numerator by \Eq{eq:completeness}.  For example, on-shell tracelessness of $\chi_{\mu\nu\rho\sigma}$ implies that, when the on-shell condition is satisfied, $\Pi_{\mu\nu\rho\sigma\alpha\beta\gamma\delta}$ vanishes when any two indices among the first set of four are contracted and similarly for the second set. Because we do not {\it a priori} know the form of the propagator numerator, we must construct it purely from its symmetries and the on-shell transversality and tracelessness conditions. 

The most general construction begins by considering $\Pi_{\mu\nu\rho\sigma\alpha\beta\gamma\delta}$ to be an arbitrary eight-index tensor built out of $\eta_{\mu\nu}$ and $k_\mu$. Then, in general $D$, we impose the requisite symmetries coming from the index properties of the Weyl tensor and symmetry on exchange of the two copies of $\chi_{\mu\nu\rho\sigma}$: antisymmetry on the first and second pairs of indices, symmetry under the exchange of the first and second index pairs, symmetry under the exchange of the first and second sets of four indices, the first Bianchi identity $\Pi_{\mu[\nu\rho\sigma]\alpha\beta\gamma\delta}
=\Pi_{\mu\nu\rho\sigma\alpha[\beta\gamma\delta]}=0$, on-shell tracelessness on each set of four indices (for arbitrary metric contraction of two indices), and on-shell transversality per \Eq{eq:transverse}. We discover that these conditions are enough to fix the propagator numerator $\Pi_{\mu\nu\rho\sigma\alpha\beta\gamma\delta}$ up to some as-yet-unspecified coefficient $\beta$:
\begin{widetext}
\be 
\begin{aligned}
\Pi^{\mu\nu\rho\sigma}_{\;\;\;\;\;\;\;\;\,\alpha\beta\gamma\delta} &= 
\beta\Big[2(D-2)(D-3)\Big(\Pi^{\mu}_{\;\;[\alpha}\Pi^{\vphantom{p}\nu}_{\;\;\beta]}\Pi^{\rho}_{\;\;[\gamma}\Pi^{\vphantom{p}\sigma}_{\;\;\delta]}+\Pi^{\mu}_{\;\;[\gamma}\Pi^{\vphantom{p}\nu}_{\;\;\delta]}\Pi^{\rho}_{\;\;[\alpha}\Pi^{\vphantom{p}\sigma}_{\;\;\beta]}\Big)\\
&\qquad \; +(D-2)(D-3)\left(\Pi^{[\mu}_{\;\;\delta\vphantom{]}}\Pi^{\vphantom{p}\nu]}_{\;\;[\alpha}\Pi^{[\rho}_{\;\;\beta]}\Pi^{\vphantom{p}\sigma]}_{\;\;\gamma\vphantom{]}}-\Pi^{[\mu}_{\;\;\gamma\vphantom{]}}\Pi^{\vphantom{p}\nu]}_{\;\;[\alpha}\Pi^{[\rho}_{\;\;\beta]}\Pi^{\vphantom{p}\sigma]}_{\;\;\delta\vphantom{]}}\right)\\
&\qquad \; -3(D-2)\Big(\Pi^{[\mu}_{\;\;[\alpha}\Pi^{\nu][\rho}_{\vphantom{()}}\Pi_{\beta][\gamma}^{\vphantom{()}}\Pi^{\sigma]}_{\;\;\delta]} + \Pi^{[\mu}_{\;\;[\gamma}\Pi^{\nu][\rho}_{\vphantom{()}}\Pi_{\delta][\alpha}^{\vphantom{()}}\Pi^{\sigma]}_{\;\;\beta]}\Big)\\
&\qquad \; +12 \Pi^{\mu[\rho}_{\vphantom{()}}\Pi^{\sigma]\nu}_{\vphantom{()}}\Pi_{\alpha[\gamma}^{\vphantom{()}}\Pi_{\delta]\beta}^{\vphantom{()}}\Big],\end{aligned}\label{eq:BigPi}
\ee
\end{widetext}
where we found that the result could be written in terms of the Proca propagator numerator
\be
\Pi_{\mu\nu} = \eta_{\mu\nu} + \frac{k_\mu k_\nu}{\mu^2}.
\ee
The appearance of this dependence on the projection operator $\Pi_{\mu\nu}$ is not surprising given the transversality condition \eqref{eq:transverse}.  However, we emphasize that we did not assume beforehand that $\Pi_{\mu\nu\rho\sigma\alpha\beta\gamma\delta}$ could be expressed as a function of the Proca propagator numerator.

Now, by the completeness relation \eqref{eq:completeness}, the full trace of the propagator numerator counts the number of physical degrees of freedom, so we must have $\Pi_{\mu\nu\rho\sigma}^{\;\;\;\;\;\;\;\;\,\mu\nu\rho\sigma}>0$. Specifically, the number of independent physical degrees of freedom in $\chi_{\mu\nu\rho\sigma}$ is just the number of possible polarizations. This is the number of tensors $\varepsilon_{i\mu\nu\rho\sigma}$ with the symmetries of the Weyl tensor that respect the transversality condition. Working through the combinatorics is straightforward and one finds that the number of physical degrees of freedom is
\be
N = \frac{1}{12}(D+1)D(D-1)(D-4). 
\ee
On the other hand, from \Eq{eq:BigPi}, we find the beautiful expression
\be 
\begin{aligned}
&\Pi_{\mu\nu\rho\sigma}^{\;\;\;\;\;\;\;\;\,\mu\nu\rho\sigma}\\&\;\;\;=2\beta(D+1)D(D-1)(D-2)(D-3)(D-4),\label{eq:traceBigPi}
\end{aligned}
\ee 
which for $D>4$ is positive if and only if $\beta>0$. Requiring that $\Pi_{\mu\nu\rho\sigma}^{\;\;\;\;\;\;\;\;\,\mu\nu\rho\sigma} = N$, we have
\be 
\beta = \frac{1}{24(D-2)(D-3)}.\label{eq:beta}
\ee
Equivalently, we recall that a propagator numerator, when taken on shell, is a projector onto the space orthogonal to $k_\mu$ \cite{WeinbergQFT} and onto tensors with the requisite index symmetries.  Requiring that the propagator numerator be idempotent as a projection operator thus fixes the normalization.

\section{Integrating Out Massive States\label{sec:integrating}}

We can now compute the higher-curvature corrections induced by integrating out $\chi$. As noted earlier, interactions between gravitons and two or more powers of $\chi$ can contribute to higher-curvature corrections given the mixing term in \Eq{eq:operator}.  Thus, to study graviton scattering at low energies, it would be necessary to do a proper accounting of all the interactions involving $\chi$ beyond even \Eq{eq:operator}.  As this is rather cumbersome, it is more convenient to compute the off-shell two-point function for the graviton.  This low-energy operator receives contributions from \Eq{eq:operator}, but crucially is independent of the interactions nonlinear in $\chi$.

Armed with a general parameterization of the couplings and spectrum of the massive states, we can now integrate them out.
Using \Eqs{eq:BigPi}{eq:beta}, one finds
\be C^{\mu\nu\rho\sigma}\Pi_{\mu\nu\rho\sigma\alpha\beta\gamma\delta}C^{\alpha\beta\gamma\delta}
\stackrel{k\rightarrow0}=C_{\mu\nu\rho\sigma}C^{\mu\nu\rho\sigma}.
\ee
Since we are computing the two-point function for gravitons, we are implicitly expanding $C_{\mu\nu\rho\sigma}$ at linear order in gravitons.
Integrating out $\chi_{\mu\nu\rho\sigma}$ at low momentum transfer, we obtain the effective operator 
\be 
\frac{y^2}{2} C_{\mu\nu\rho\sigma}C^{\mu\nu\rho\sigma} \int_0^\infty\frac{{\rm d}\mu^{2}}{\mu^2}\rho(\mu^2).
\ee
We then deduce the coefficient of the Weyl-squared operator in \Eq{eq:actionWeyl},
\be
\lambda = \frac{y^2}{2} \int_0^\infty\frac{{\rm d}\mu^{2}}{\mu^2}\rho(\mu^2)\geq 0.
\ee
Thus, since the spectral function is nonnegative by unitarity, the sign of the coefficient $\lambda$ of the GB operator is nonnegative in a consistent tree-level ultraviolet completion in $D>4$.

This bound is consistent with results from string theory \cite{GrossWitten,Zwiebach,GrossSloan,Tseytlin}. Moreover, our bound constitutes a requisite consistency condition for any candidate tree-level theory of quantum gravity. Proving positivity of the GB coefficient using a different approach---analytic dispersion relations---is the subject of current ongoing research \cite{Reece}, though subtleties exist in applying analyticity bounds to graviton amplitudes \cite{IRUV,GravityBounds}. While standard axioms of quantum field theory, e.g., locality, may be violated in quantum gravity, dispersion relations themselves seem to remain robust \cite{Giddings:2009gj}.

Delineating the boundary between the swampland and the landscape can provide insights for model-building and for our broader understanding of gravitational ultraviolet completion of quantum field theories. Open problems include finding ways to apply infrared consistency bounds in nonperturbative contexts, as well as connecting bounds obtained from infrared- and ultraviolet-dependent reasoning.

\medskip

\noindent {\it Acknowledgments}: We thank Brando Bellazzini and Matt Reece for useful discussions and comments. C.C. is supported by a Sloan Research Fellowship and a DOE Early Career Award under Grant No. DE-SC0010255. G.N.R.~is supported by a Hertz Graduate Fellowship and a NSF Graduate Research Fellowship under Grant No.~DGE-1144469.

\bibliographystyle{utphys}
 
\bibliography{CurvatureSquared}

\end{document}